\documentclass[conference,a4paper]{IEEEtran}
\IEEEoverridecommandlockouts
\usepackage{graphicx}
\usepackage{amsmath,amssymb,dsfont,bbm,epstopdf,pgfplots} 
\usepackage{amstext,amsfonts,amssymb,graphicx,stmaryrd}
\usepackage{color}
\usepackage{cite}
\usepackage{graphics}
\usepackage{tikz}
\usepackage{pgfplots}
\usepackage{stfloats}
\usetikzlibrary{shapes, arrows, decorations.markings, arrows.meta}
\usetikzlibrary{patterns} 
\allowdisplaybreaks
\graphicspath{{.}{./Figures/}} 
\allowdisplaybreaks[4] 
\usepackage{amscd}
\usepackage{epsfig}
\usepackage{graphics}
\usepackage{psfrag}
\usepackage{rotating}
\usepackage{amsmath}
\usepackage{amsfonts}
\usepackage{url}
\usepackage{color}
\usepackage[caption=false,font=footnotesize]{subfig}
\usepackage{epstopdf}
\usepackage{amsthm}
\usepackage{tikz}
\usepackage{mathtools}
\usetikzlibrary{shapes.geometric,arrows,shapes,decorations,backgrounds}
\usepackage[final]{pdfpages}
\newtheorem{theorem}{Theorem}

\newtheorem{remark}{Remark}
\newtheorem{definition}{Definition}

\newtheorem{corollary}[theorem]{Corollary}

\newfont{\bbb}{msbm10 scaled 500}

\newfont{\bb}{msbm10 scaled 1100}

\newcommand{\Mc}{{\cal M}}

\newcommand{\Sc}{{\cal S}}

\newcommand{\Xc}{{\cal X}}
\newcommand{\Yc}{{\cal Y}}
\newcommand{\Zc}{{\cal Z}}

\DeclareFontFamily{U}{cmfi}{}
\DeclareFontShape{U}{cmfi}{m}{n}{ <-> cmfi10 }{}
\DeclareSymbolFont{CMFI}{U}{cmfi}{m}{n}

\newcommand{\define}{\stackrel{\triangle}{=}}

\newcommand{\D}{\mathsf{D}}
\newcommand{\R}{\mathsf{R}}

\usepackage{multicol}
\begin{document}
	
	\title{Integrated  Communication and Receiver Sensing with Security Constraints  on Message and State }
	
		\author{
		\IEEEauthorblockN{Mehrasa Ahmadipour\IEEEauthorrefmark{1}, Mich\`ele Wigger\IEEEauthorrefmark{2}, Shlomo Shamai\IEEEauthorrefmark{3} 
		}
		\IEEEauthorblockA{\small\IEEEauthorrefmark{1}
			UMPA, ENS de Lyon
			\url{mehrasa.ahmadipour@ens-lyon.fr}}
		\IEEEauthorblockA{\small\IEEEauthorrefmark{1} LTCI Telecom Paris, IP Paris, 91120 Palaiseau, France, Emails:
			\url{michele.wigger@telecom-paris.fr}}
		\IEEEauthorblockA{\small\IEEEauthorrefmark{2} ,  Email: sshlomo@ee.technion.ac.il
		}
	}
	
	
	
	\maketitle

	
	\IEEEpeerreviewmaketitle

	\begin{abstract}
		We study the state-dependent wiretap channel with non-causal channel state informations at the encoder in an integrated sensing and communications (ISAC) scenario. 
		In this scenario,  the transmitter  communicates a message and a state sequence to a legitimate receiver while keeping the message and state-information secret from an external eavesdropper. This paper presents a new achievability result for this doubly-secret scenario, which recovers as special cases the best-known  achievability results for the setups  without security constraints or with only a security constraint on the message. The impact of the secrecy constraint (no secrecy-constraint, secrecy constraint only on the message, or on the message and the state) is analyzed at hand of a Gaussian-state and Gaussian-channel example. 
	\end{abstract}
	\section{Introduction}
	
	Great scientic and technological efforts are currently being made to eifficiently integrate sensing and communication (ISAC) into a common hardware and  bandwidth \cite{YaoCorreState2022,LiuSurvey2022,Tong,ISAC_Liu,zheng2019radar,dokhanchi2019adaptive}. This trend is driven by spectrum scarcity, the enlargement of the communication spectrum closer to the traditional radar spectrum, the conceptual similarity between the two tasks (emitting specific  waveforms and detecting parameters based on  received signals), as well as economic pressure to reduce hardware costs.  Radar systems can be divided into two families:  \emph{mono-static radar}  where the same terminal emits the waveform and senses the environment based on the backscattered signal and \emph{bi-static radar} where  the  sensing terminal exploit the scattered signals emitted by other terminals. Information-theoretic works have considered both types of systems, where in the literature on  mono-static radar \cite{kobayashi2018joint, ISAC_BC,ISAC_MAC}   the radar receivers  typically have to reconstruct the channel's state sequence with smallest possible distortion , while in the literature on   bi-static radar \cite{Joudeh2022Discrim,Joudeh2021JBinaryDetect,Bloch2022,ISAC_strong} they aim to determine an underlying binary (or multi-valued) parameter with largest possible error exponent. This latter scenario has even been investigated for classical-quantum channels \cite{Uzi_ISAC}.
	
\begin{figure}[t!]
		\centering
		\includegraphics[width=0.46\textwidth]{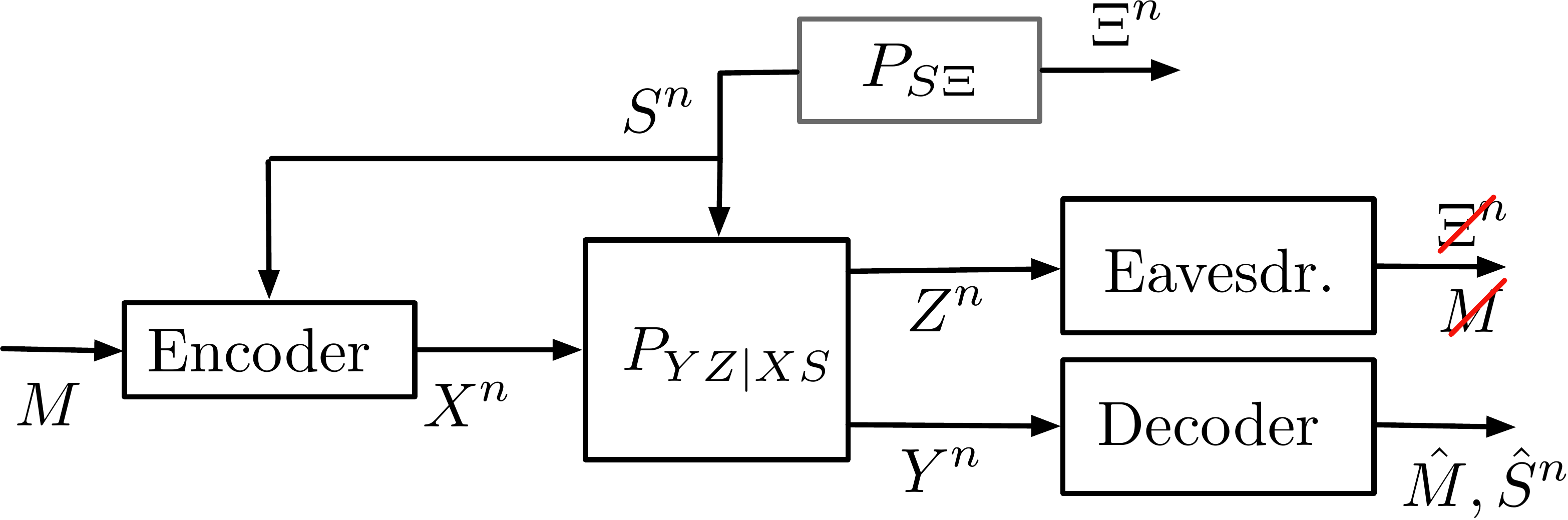}
		\caption{ State-dependent wiretap channel with non-causal  channel state information at the transmitter and security constraints on the message and the state.}
		\label{fig:model}
		\vspace{-5mm}
	\end{figure}

	In this work we consider an ISAC problem with bi-static sensing, where the sensing terminal coincides with the receiver of the data communication, see Figure~\ref{fig:model}. The receiver thus not only decodes the transmitted signal but also reconstructs the channel state-sequence up to a given distortion. We assume that the transmitter knows this state-sequence $S^n$ perfectly and in advance (as in the famous Gel'fand-Pinsker  \cite{GelfandPinsker} or dirty-paper setups \cite{DPC_Costa}) and can thus actively help the receiver in his estimation. Bounds on the optimal rate-distortion tradeoff of the described setup has been derived  in \cite{Sutivong2005,zhang2011joint, Choudhuri2012noncausal}. In the present work, we impose  the additional security constraints that  the  transmitted message and part of the state $S^n$ have to be kept secret from an external eavesdropper. In this sense, our model is  an extension of the models in \cite{Sutivong2005,zhang2011joint, Choudhuri2012noncausal} but with an external eavesdropper that is not allowed to learn neither the message nor the state. It can also be considered an extension of the wiretap channel  with non-causal state-information \cite{SD_WT} to include the sensing performance and the secrecy constraint on the state. The intriguing feature in our model is that the transmitter should describe the state to the legitimate receiver, but mask  \cite{Merhav2007} it from the eavesdropper. 
		

Related to this work is also  \cite{Gunlu}, which however considers a mono-static radar system, and \cite{Bunin2020} which considers  a state-dependent wiretap channel (without sensing) where   the transmitter additionally has to extract a key and  convey this to the receiver while keeping it secret from the eavesdropper. 

In this paper we establish an achievable rate-distortion region  for the state-dependent wiretap channel with receiver state-sensing and where besides the message also parts of the state have to kept secret from an external eavesdropper. 
		Our result is built on a two-layer superposition likelihooed encoding scheme, similar to   \cite{SD_WT}. It includes as special cases the results we mentioned in above paragraph. At hand of a Gaussian example, we  numerically characterize the penalty on the rate-distortion tradeoff caused by the security constraints on the message and on the state.

	\section{System Model and Main Result}\label{sec:setup}
	
	Formally, the model consists of the following elements.
	\begin{itemize}
		\item  A memoryless state sequence $\{(S_i)\}_{i\geq 1}$ whose samples at any given time $i$ are distributed according to a given law $P_{S}$ over the finite state alphabet $\Sc$.
		\item Given that at time-$i$ the Tx sends input $X_{i}=x$  and given  state realization $S_{i}=s_{i}$,  the Rx observes the  time-$i$ output $Y_{i}$  and the eavesdropper observes signal $Z_{i}$ distributed according to the stationary channel transition law $ P_{YZ|SX}(\cdot,\cdot|s,x)$,
		\item Input and output 
		alphabets $\Xc , \Yc,  \Zc, \Sc$ are assumed  finite. 
	\end{itemize}  
	
	A rate-$R_M$ and blocklength-$n$ code  consists  of:
	\begin{enumerate}
		\item A message set $\Mc_n\define\{1,2,\cdots, 2^{nR_M}\}$;
		\item An encoder that assigns a symbol $x_i (m, s ) \in X$
		to each message $m\in \Mc_n$ and each state
		sequence $s^n\in \Sc^n$; 
		\item  A decoder that assigns a message estimate $\hat{m}$ and a state sequence estimate $\hat{s}^n\in {\hat{\Sc}^n} $ to each received sequence $y^n\in \Yc^n$ where  $\hat{\Sc}$ is a given  reconstruction alphabet.
		\item An encoding function: $f_n:\Mc_n\times \Sc^n\to \Xc^n$
		\item A decoding function: $\phi_n: \Yc^n\to \Mc_n$
		\item  A state estimator  $\psi_n \colon \Yc^n \to \hat{\Sc}^n$.
	\end{enumerate}
	
	
	We assume that $M$ is uniformly distributed over the
	message set so $P_{M}(m)=\frac{1}{2^{nR_M}}$. 
	The probability of decoding error is defined as:
	\begin{IEEEeqnarray}{rCl}
		P^{(n)}_e& := &\textnormal{Pr}\Big( \hat{M}\neq M \Big).
	\end{IEEEeqnarray}
	The fidelity of the state estimate at the Rx is measured by the expected distortion
	\begin{equation}
		\Delta^{(n)}:= \frac{1}{n} \sum_{i=1}^n \mathbb{E}[d_{\textnormal{R}}(S_{ i},  \hat{S}_{ i})], 
	\end{equation}
	for a given bounded distortion function $d_{\textnormal{R}}(\cdot,\cdot)$.
	
	The eavesdropper should not be able to learn a random sequence $\Xi^n$ that is obtained by passing the  state sequence $S^n$ through a memoryless channel $P_{\Xi|S}$ independent of the message and the communication channel. 
	The information leakage to the eavesdropper 
	is   measured by the mutual information
	\begin{equation}\label{InfLeak1}
		l_s^{(n)}\triangleq   I(M,\Xi^n;Z^n). 
	\end{equation}
	
	
	\begin{definition} In the described setup, a rate-distortion pair $(\R_M,  \D)$ is called securely-achievable if there exists  a sequence (in $n$) of  rate-$R_M$ and blocklength $n$-codes that simultaneously satisfy the three asymptotic constraints:
		\begin{subequations}\label{eq:asymptotics}
			\begin{IEEEeqnarray}{rCl}
				\lim_{n\to \infty}	P^{(n)}_e 
				&=&0 \\
				\lim_{n\to \infty}	l_s^{(n)} &=&0 \label{eq:asymptotic_sec} \\
				\varlimsup_{n\to \infty}	\Delta^{(n)}& \leq& \D.\label{eq:asymptotics_dis}
				%
			\end{IEEEeqnarray}
		\end{subequations}
	\end{definition}
	
	\begin{remark} 
		Notice that Condition~\eqref{eq:asymptotic_sec} is equivalent to requiring that the sum of  mutual informations $I(M;Z^n)+ I(\Xi^n;Z^n)$ vanishes asymptotically as $n\to \infty$.
	\end{remark} 

	\begin{theorem}\label{thm:ach_equivocation}
		For any pmf $P_{UVX\mid S}$ so that for the associated tuple $(S,\Xi, U, V, X,Y Z)\sim P_{S}P_{\Xi|S}P_{UVX\mid S} P_{YZ|XS}$, the random variable $\Xi$ is independent of the pair $(U,Z)$,
		\begin{equation}
			\Xi \perp (U,Z)
		\end{equation} and any function $g(\cdot)$ on appropriate domains, all pairs  
		$(R_M, D)$ satisfying the following  inequalities
		\begin{IEEEeqnarray}{rCl}\label{eq:thm:ach_equivocation}
			R_M&\leq& I(U,V;Y)-I(U,V;S) \label{eq:e1} 
			\\
			R_M&\leq& I(V;Y\mid U)-	I(V;\Xi,Z\mid U) 
			\nonumber\\
			&&\hspace{2cm}+\min\{0, I(U;Y)-I(U;S)\}\label{eq:e2}
			\label{eq2:Th1}	
		\end{IEEEeqnarray}
		and
		\begin{IEEEeqnarray}{rCl}
			E[d(S,g(U,V,Y))]&\leq& D,
		\end{IEEEeqnarray}
		are securely achievable. 
	\end{theorem}

	Notice that when the entire state $S$ has to be kept secret, $\Xi=S$, then $U$ has to be chosen independently of $S$ and thus $I(U;S)=0$ and the minimum in the right-hand side of \eqref{eq:e2} evaluates to 0. Moreover, for $\Xi=S$ the right-hand side of \eqref{eq:e1}  is larger  than the right-hand side of \eqref{eq:e2} because $I(V;S,Z\mid U)\geq I(V;S\mid U)$. Thus, for $\nu(S)=S$, Constraint~\eqref{eq:e1} is less stringent than Constraint~\eqref{eq:e2} and we obtain the following corollary, after observing that $U$ only plays the role of a convexification random variable. 
	\begin{corollary}[Fully-Secret State]\label{cor:secret_state}
		Assume $\Xi=S$. Then the convex hull of all rate-distortion pairs  
		$(R_M, D)$ is achievable that
		satisfy the constraints
		\begin{IEEEeqnarray}{rCl}
			R_M&\leq& I(V;Y)-	I(V;S,Z)		\end{IEEEeqnarray}
		and
		\begin{IEEEeqnarray}{rCl}
			E[d(S,g(V,Y))]&\leq& D,
		\end{IEEEeqnarray}	
		for some  function $g(\cdot)$ on appropriate domains and  pmf $P_{VX\mid S}$ where the associated tuple $(S,\Xi, V, X,Y Z)\sim P_{S}P_{\Xi|S}P_{VX\mid S} P_{YZ|XS}$ has 	$S$ independent of $Z$, 
		\begin{equation}
			S\perp Z.
		\end{equation}
		
	\end{corollary}
	
	On the other extreme, we might only wish to keep the message secret but not the state, i.e. $\Xi=$const. As shown in Section~\ref{sec:simplification},  in this case Theorem~\ref{thm:ach_equivocation}  simplifies  to:
	\begin{corollary}[No Secrecy Constraint on  State]\label{cor:no_secret_state}
		Assume $\Xi=$const. For any pmf $P_{UVX\mid S}$ and any function $g(\cdot)$ on appropriate domains, all pairs  
		$(R_M, D)$ satisfying
		\begin{IEEEeqnarray}{rCl}
			R_M&\leq& \min\{ I(U,V;Y)-I(U,V;S) ,
			\nonumber\\
			&&\hspace{2cm} \; I(V;Y\mid U) -	I(V;Z\mid U)\}
		\end{IEEEeqnarray}
		and
		\begin{IEEEeqnarray}{rCl}
			E[d(S,g(U,V,Y))]&\leq& D,
		\end{IEEEeqnarray}
		are securely achievable   rate-distortion pairs. It suffices to consider pmfs  $P_{UVX\mid S}$ so that $I(U;Y) \geq I(U;S)$.
	\end{corollary}
	Notice that for sufficiently large distortion constraints $D$, Corollary~\ref{cor:no_secret_state} recovers the achievability result for the state-dependent wiretap channel with non-causal state-information at the encoder \cite[Theorem~1]{SD_WT}. 
	
	Finally, we consider the special case where $Z$ is independent of the input-state pair $(X,S)$, which corresponds to the setup without secrecy constraint studied in \cite{choudhuri2012}. In this special case, Theorem~\ref{thm:ach_equivocation} can be simplified by choosing $U=$const., in which case Corollary~\ref{cor:no_secret_state} evaluates to: 
	\begin{corollary}[No Eavesdropper, coincides with Theorem~1 in \cite{choudhuri2012}]\label{cor:no_eve}
		Assume that $Z$ is independent of the input-state pair $(X,S)$. Then, for any pmf $P_{VX\mid S}$ and any function $g(\cdot)$ on appropriate domains, all non-negative rate-distortion pairs  
		$(R_M, D)$ satisfying
		\begin{IEEEeqnarray}{rCl}
			R_M&\leq&  I(V;Y)-I(V;S)
		\end{IEEEeqnarray}
		and
		\begin{IEEEeqnarray}{rCl}
			E[d(S,g(V,Y))]&\leq& D,
		\end{IEEEeqnarray}
		are  achievable. 
	\end{corollary}
	
	Comparing Corollary \ref{cor:secret_state}  where both message and state have to be kept secret with Corollary \ref{cor:no_eve} where no secrecy constraint is applied, we see that the price for achievable double state-and-message secrecy in our scheme is that $S$ has to be independent of $Z$ and that the rate has to be reduced by the mutual information quantity $I(V;Z|S)=I(SV;Z)$. 
	
	\subsection{A Gaussian Example}
	Consider Gaussian channels both to the legitimate receiver
	\begin{IEEEeqnarray}{rCl}\label{ex:Gaussian1user}
		Y_i &= &X_i +S_i + N_{i} ,
	\end{IEEEeqnarray}
	and to the eavesdropper 
	\begin{IEEEeqnarray}{rCl}
		Z_i &= &aX_i +bS_i + N_{e,i},
	\end{IEEEeqnarray}
	for given parameters $a=0.7, b=0.3$,  where $\{S_i\}$ is an independently and identically distributed  (i.i.d.) zero-mean Gaussian of power $Q=3$ and the noise sequences are also i.i.d. zero-mean Gaussian of variances $\sigma^2=1$ and $\sigma_{e}^2=4$ and independent of each other and of the inputs and the states.  The channel inputs are average blockpower constrained to power $P=30$. Also, assume that 
	\begin{equation}
		\Xi = S+A,
	\end{equation}
	for $A$ a zero-mean Gaussian random variable independent of all other random variables and of variance $\sigma_A^2\geq 0$. This setup covers the scenario where the entire state-sequence has to be kept secret, with the choice $\sigma_A^2=0$. and (with a slight abuse of notation) the scenario where the state does not have be be kept secret at all, with the choice $\sigma_A^2 \to \infty$.
	
	We use the squared error  $d(s,\hat{s})=(s-\hat{s})^2$ to measure the distortion at the receiver.
	
	We shall numerically compare the achievability results in Corollaries~\ref{cor:secret_state}, \ref{cor:no_secret_state}, \ref{cor:no_eve} for this Gaussian example, to quantify the rate-penalty imposed by the various secrecy-constraints. To this end, we choose the following Gaussian auxiliaries: 
	\begin{subequations}\label{eq:auxiliaries}
	\begin{IEEEeqnarray}{rCl}
		U &= & F+ \delta S+G\\
		V & =& T+ \alpha S+G\\
		X & = & T +F+\epsilon G+\gamma S,
	\end{IEEEeqnarray}
	\end{subequations}
	where $T$, $F$, and $G$  Gaussian random variable independent each of other and of the state $S$, and of variances $\sigma_T^2, \sigma_F^2\geq 0$ so that $\sigma_T^2+\sigma_F^2+ \epsilon^2 \sigma_G^2+ \gamma^2 Q \leq P$.
	Notice that for $\Xi=S$ we are obliged to choose $\gamma=-\frac{b}{a}$ to ensure that $Z$ is independent of $S$.  In this case, we can also choose $U$  to be constant (i.e., $F,G$ constants and $\delta=0$), see Corollary~\ref{cor:secret_state}.

\begin{figure}
		\begin{center}
			\centering
	\includegraphics[scale=0.8]{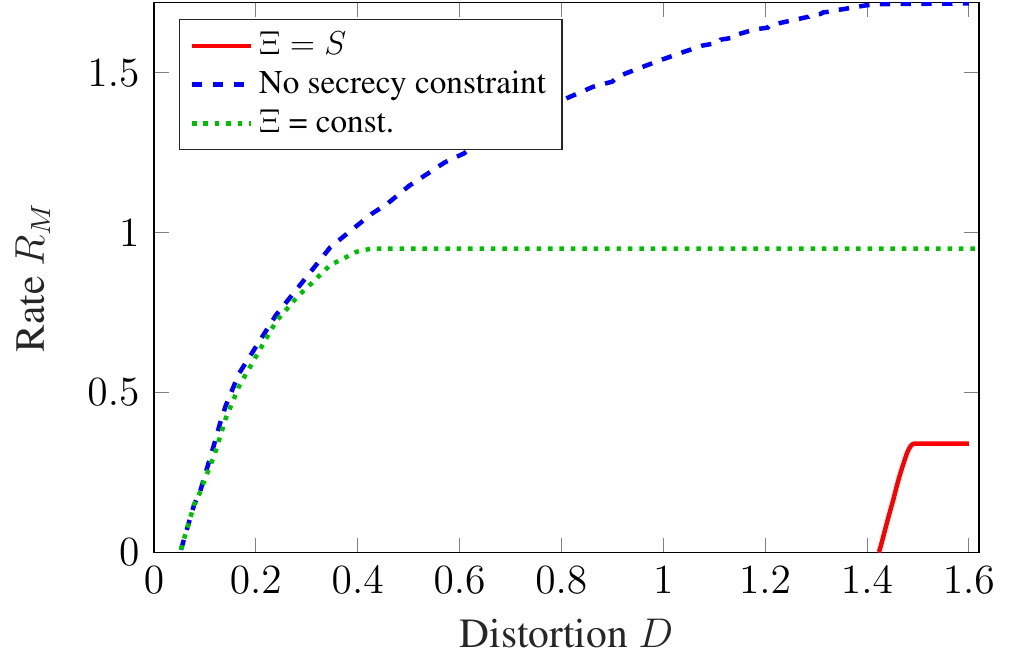}%
	\caption{Comparison of the achievable rate-distortion tradeoffs for a Gaussian channel with security constraints on both the message and the state and no security constraints at all.}
	\label{fig:plot}
\end{center}
		\end{figure}

	Figure~\ref{fig:plot} illustrates the distortion-rate tradeoffs achieved  by our Corollaries~\ref{cor:secret_state}, \ref{cor:no_secret_state}, \ref{cor:no_eve}  for the described choice of auxiliaries. Notice that without any secrecy constraint, in the Gaussian case the achievability result in Corollary~\ref{cor:no_eve} is tight, as shown in \cite{Sutivong2005}. The largest achievable rate equals the dirty-paper capacity \cite{DPC_Costa} of the channel to the legitimate receiver $C=1/2\log(1+P/\sigma_N^2)=1.717$. In this no-secrecy setup the minimum distortion is achieved by simply sending a scaled version of the channel and equals 
	\begin{equation}
		D_{\min, \textnormal{no-secrecy}}=Q \frac{ \sigma_N^2}{(\sqrt{Q} + \sqrt{P})^2+\sigma_N^2}=0.056.
	\end{equation}
	Trivially, the same minimum distortion is also achievable in the classical wiretap setup where only the message but not the state have to be kept secret.
	
	Finding the minimum distortion under a secrecy constraint on the state is more challenging, the same as finding the maximum rate when the message has to be kept secret (both in the scenarios with and without secrecy constraint on the state). We can however easily deduce that the minimum distortion under a secrecy constraint on the state cannot be achieved by an uncoded strategy. In fact, sending $X^n=-\frac{b}{a} S^n$ without any additional codeword (recall that the transmit signal $X$ has to subtract the term  $\frac{b}{a} S^n$ so as to keep the state $S^n$ secret from the eavesdropper) achieves distortion
	\begin{equation}
		D_{\min, \textnormal{uncoded,secret}}=Q \frac{ \sigma_N^2}{Q\left(1- \frac{b}{a}\right)^2+\sigma_N^2}=1.516,
	\end{equation}
	which is strictly larger than the minimum distortion $1.423$ achieved by Theorem~\ref{thm:ach_equivocation} (see the red line in Figure~\ref{fig:plot}) under the fully-secret state criterion $\Xi=S$ and with the set of auxiliaries in \eqref{eq:auxiliaries}.

	\section{Proof of Theorem \ref{thm:ach_equivocation}}

	Choose a conditional distribution $P_{XUV|S}$ over $\mathcal{X}\times \mathcal{U} \times \mathcal{V}$ for auxiliary alphabets $\mathcal{U}$ and $\mathcal{V}$, and a  reconstruction function $g\colon \mathcal{U} \times \mathcal{V}\times  \mathcal{Y} \to \mathbb{R}_0^+$ so that for the tuple $(S,\Xi,U,V,X, Y,Z) \sim P_{S}P_{\Xi|S}P_{XUV|S} P_{YZ|XS}$ the random variable $\Xi$ is independent of the pair $(U,Z)$: 
	\begin{IEEEeqnarray}{rCl}\label{eq:independent}
		\Xi \perp (U,Z)
	\end{IEEEeqnarray} and the distortion constraint is satisfied: 
	\begin{equation}\label{eq:disto}
		\mathbb{E}[ d_{\textnormal{R}}(S, g(U,V,Y)) ]\leq D_{\textnormal{R}}.
	\end{equation}
	Fix a small number $\epsilon>0$ and a large blocklength $n$. 	
	Pick auxiliary rates  $R_{I}$,  and $R_J$ satisfying
	\begin{subequations}\label{eq:conditions}
		\begin{IEEEeqnarray}{rCl}
			R_I  & \geq  & I(U;S )\\
			R_I+R_J & \geq &  I(UV;S)\\
			R_J & \geq & 	I(V;\Xi,Z|U)
		\end{IEEEeqnarray}
	\end{subequations}

	\subsection{Code Construction}
	Construct a superposition code as follows.
	\begin{itemize}
		\item A lower-level code $\mathcal{C}_{U}$ consisting of $2^{n {R}_{I}}$ codewords $\{u^n(i)\}$  is constructed by drawing all entries i.i.d. according to the marginal pmf $P_{U}$ of 
		\begin{IEEEeqnarray}{rCl}\label{eq:joint_pmf}
			P_{SUVXYZ}=	P_{S}P_{XUV|S} P_{YZ|XS}. 
		\end{IEEEeqnarray} 
		\item An upper-level code $\mathcal{C}_{V}(i)$ consisting of $2^{n( {R}_{J}+R_M)}$  codewords $\{v^n(m,j\mid i)\}$ is constructed for each $i\in [ 2^{n {R}_{I}}]$, by drawing the $t$-th entry  of each codeword according to $P_{V|U}(\cdot \mid u_t(i))$ where  $u_t(i)$ denotes the $t$-th entry of  codeword  $u^n(i)$. 
	\end{itemize}  
	The realization of the codebook is revealed to all parties.
	
	\subsection{Encoding and Decoding}
	
	The transmitter applies a likelihood encoder. That means, based on the state-sequence $S^n=s^n$ that it observes  and the message $M=m$ that it wishes to convey, it randomly picks the indices $(I^*,J^*)$ according to the conditional pmf
	\begin{IEEEeqnarray}{rCl}
	&&	P_{\textnormal{LE}}(i,j \mid m,s^n) 
	\nonumber
	\\
	&&= \frac{ P_{S|UV}(s^n| u^n(i), v^n(m,j\mid i)) }{ \underset{i\in  [2^{n {R}_{I}}]}{\sum} \underset{j\in[2^{n {R}_{J}}]}{\sum} P_{S|UV}(s^n| u^n(i), v^n(m,j\mid i) )}.
	\end{IEEEeqnarray}
	It then generates the random input sequence $X^n$ by passing the pair  of codewords $u^n(I^*)$ and $v^n (m,J^*\mid I^*)$ and the state sequence $s^n$ through the memoryless channel $P_{X|UVS}$.

	The receiver observes the channel outputs $Y^n=y^n$ and looks for a triple $(\hat i, \hat j, \hat m)$  so that 
	\begin{equation}
		( u^n(\hat{i}), \ v^n (\hat{m},\hat{j}| \hat{i}), \ y^n) \in \mathcal{T}_{\epsilon}^{(n)}(P_{UVY}).
	\end{equation}
	It randomly picks one of these triples and sets $(\hat{I},\hat{J}, \hat{M})=(\hat{i},\hat{j},\hat{m})$. Then it declares $\hat{M}$ as the transmitted message, and produces the state reconstruction sequence 
	\begin{equation}\label{eq:rec}
		\hat{S}^n = g^{\otimes n}(  u^n(\hat{I}), \ v^n (\hat{M},\hat{J} \mid   \hat{I}), \ y^n).
	\end{equation}
	If no triple was found, the receiver  declares an error.

	\subsection{Analysis}
	
	\subsubsection{Distortion and Error Probability}
\begin{figure*}[t]
		\centering
		\includegraphics[width=0.78\textwidth]{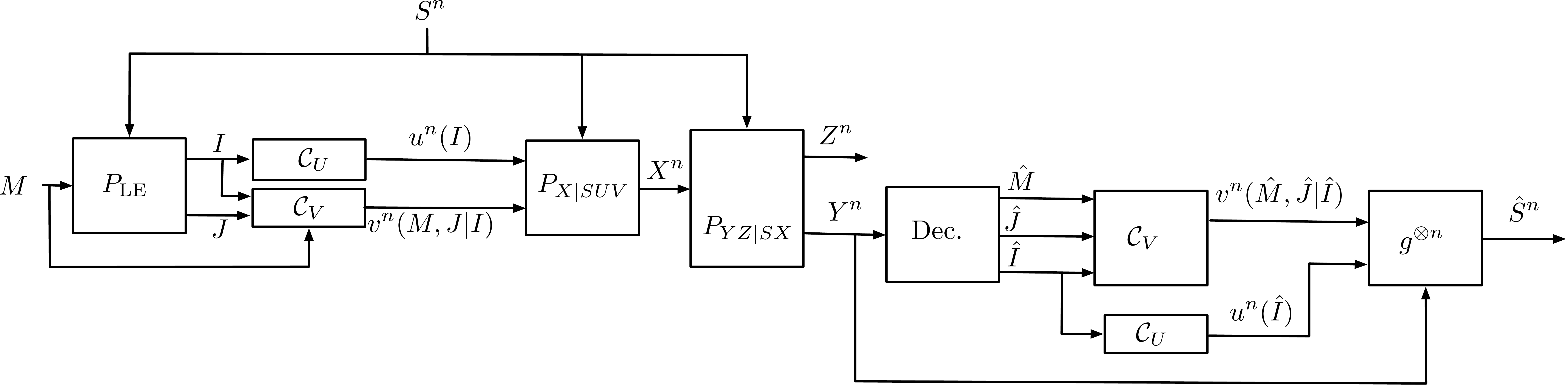}
		\caption{Real system inducing distribution $P_{MIJU^nV^nS^nX^nY^nZ^n\hat{S}^n}$.}
		\label{real}
			\vspace{2mm}
		\hrule
	\end{figure*}

	We consider two related experiments. To this end, notice that the encoding procedure is described in Figure~\ref{real}. 
	
\begin{figure*}[t]
		\centering
		\includegraphics[width=0.8\textwidth]{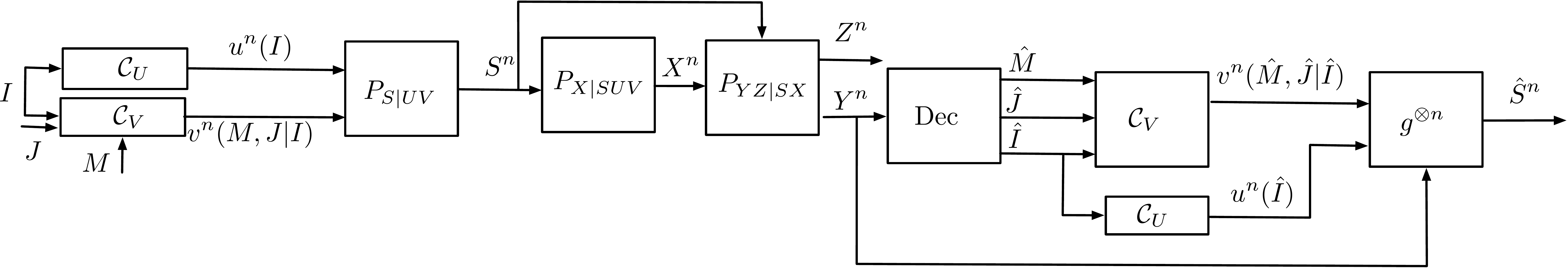}
		\caption{Auxiliary System inducing distribution $\tilde{P}_{MIJU^nV^nS^nX^nY^nZ^n\hat{S}^n}$.}
		\label{error_free}
			\vspace{2mm}
		\hrule
	\end{figure*}

	We shall prove that the joint pmf $P_{MIJU^nV^nS^nX^n Y^n Z^n}$ in this model is close in variational distance to an auxiliary distribution $\tilde{P}_{MIJU^nV^nS^nX^n Y^n Z^n\hat{S}^n}$ implied by the diagramme in Fig.~\ref{error_free}, where the the indices $I$ and $J$ are assumed uniform over the index sets $\{1,\ldots, 2^{nR_I}\}$ and $\{1,\ldots, 2^{nR_J}\}$, independent of each other and of the state sequence $S^n$ and message $M$.
	
	Notice that  for each code construction the joint pmfs ${P}_{MIJU^nV^nS^nX^nY^nZ^n\hat{S}^n}$ and $\tilde{P}_{MIJU^nV^nS^nX^n Y^n Z^n\hat{S}^n}$ factorize in a similar way: 
	\begin{IEEEeqnarray}{rCl}	
	&&	{P}_{MIJU^nV^nS^nX^n Y^n Z^n\hat{S}^n}(m,i,j,u^n,v^n,s^nx^n,y^n,z^n,\hat{s}^n) 
	\nonumber\\
	&&= \frac{1}{2^{nR}}  P^{\otimes n}_{S}(s^n) P_{\textnormal{LE}}(i,j\mid  m, s^n) \mathbbm{1}\{{u}^n(i)=u^n\}
		\nonumber \\[1.2ex]
		& & 
		\cdot \mathbbm{1}\{{v}^n(m,j\mid i)=v^n\} \cdot  P^{\otimes n}_{XYZ|UVS}(x^n,y^n,z^n|u^n,v^n,s^n)  
		\nonumber \\[1.2ex]
		& &  \hspace{4cm} \mathbbm{1}\{\hat{s}^n=g^{\otimes n} ( u^n,v^n,y^n) \} 
	\end{IEEEeqnarray}
	where $P_{\textnormal{LE}}$ denotes the conditional marginal distribution induced by the likelihood encoder, and 
	\begin{IEEEeqnarray}{rCl}	
	&&	\tilde{P}_{MIJU^nV^nS^nX^n Y^n Z^n}(m,i,j,u^n,v^n,s^n, x^n,y^n,z^n) = \nonumber\\
	&&\frac{1}{2^{n(R_M+R_I+R_J)}}  \mathbbm{1}\{{u}^n(i)=u^n\}  \mathbbm{1}\{{v}^n(m,j\mid i)=v^n\} \nonumber \\[1.2ex]
		& & \cdot  P^{\otimes n}_{S|UV}(s^n|u^n,v^n)  P^{\otimes n}_{XYZ|UVS}(x^n,y^n,z^n\mid u^n,v^n,s^n)\nonumber \\
		&& 
		\hspace{3.5cm}\cdot \mathbbm{1}\{\hat{s}^n=g^{\otimes n} ( u^n,v^n,y^n) \} .	\end{IEEEeqnarray}
	%
	%
	
	 Since  in the auxiliary system, codewords $u^n$ and $v^n$ are chosen uniformly at random over the codebooks, by standard typicality arguments, if 
	\begin{subequations}\label{eq:constraints}
		\begin{IEEEeqnarray}{rCl}	
			R_I +R_J +R_M  & < &  I(U,V;Y)\\
			R_J +R_M  & < &  I(V;Y|U).
		\end{IEEEeqnarray}
	\end{subequations} the expected (over the random choice of the codebooks) probability of wrongly decoding indices $(M,  I ,J)$ tends to 0 as $n\to \infty$. In fact, for any $M=m$, we have that 
	\begin{IEEEeqnarray}{rCl}
		\mathbb{E}\left[	p(\textnormal{error}|M=m)) \right] \to 0 \qquad \textnormal{as } n\to \infty,
	\end{IEEEeqnarray}
	where expectation is with respect to the random codebook and 
	\begin{equation}
		p(\textnormal{error}|M=m):= \Pr\left[   \hat{I} =I  \textnormal{ or }  \hat{J}=J \Big|M=m\right].
	\end{equation}
	
\begin{figure*}[!h]
		\centering
		\includegraphics[width=0.56\textwidth]{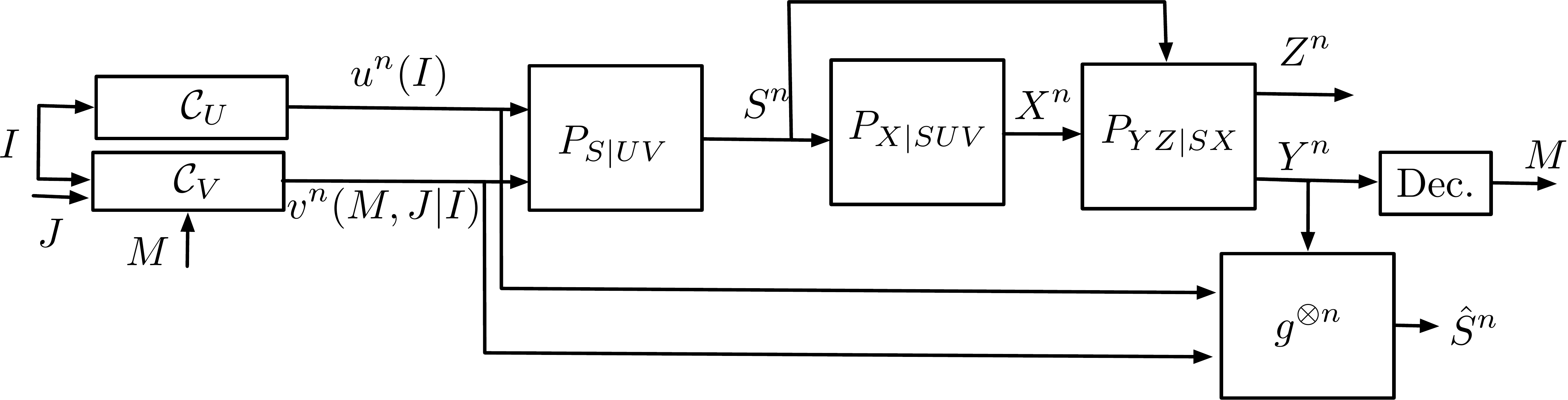}
		\caption{Ideal System inducing distribution $Q_{MIJU^nV^nS^nX^nY^nZ^n\hat{S}^n}$.}
		\label{ideal}
		\vspace{2mm}
		\hrule
			\vspace{-3mm}
	\end{figure*}

	We now introduce the ideal system in Figure~\ref{ideal}. 	
	It only differs from the system in Figure~\ref{error_free} in that error-free decoding is assumed. Notice that since under $\tilde{P}$ the decoding error probability vanishes as $n\to\infty$, we have on expectation over the random code construction: 
	\begin{IEEEeqnarray}{rCl}
	&&	\mathbb{E}\Big[ \Big\| Q_{MIJU^nV^nS^nX^n Y^n Z^n\hat{S}^n} \nonumber\\
		&&
		\hspace{2cm}-\tilde{P}_{MIJU^nV^nS^nX^n Y^n Z^n\hat{S}^n} \Big\|_1 \Big ] \to 0, 
	\end{IEEEeqnarray}
	where expectation is over the random choice of the codebooks. As a consequence, by the  boundedness of the distortion function, we have: 
	\begin{IEEEeqnarray}{rCl}\label{eq:distortionc}
		\lim_{n\to\infty} \mathbb{E}\left[		\Delta^{(n)} 	\right]  
= d_{\textnormal{R}}(S, g(U, V, Y) )  
		& \leq &D_{\textnormal{R}},
	\end{IEEEeqnarray}
	where  $(U,V, S,Y)\sim P_{UV}P_{S|UV}P_{Y|UVS}$ and the last two equalities hold by the construction of the auxiliary system and by the choice of the auxiliary random variables. 
	
	\medskip
	
	\subsubsection{Secrecy Analysis}
\begin{figure}[!h]
		\centering
		\includegraphics[width=0.35\textwidth]{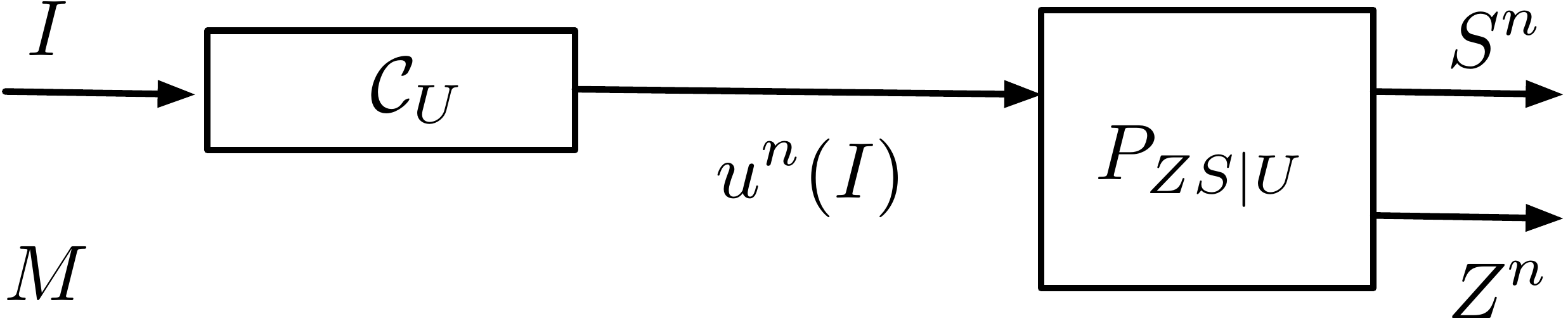}
		\caption{Ideal system used for secrecy analysis, inducing joint pmf $\bar{Q}_{MIU^n\xi(S^n)Z^n}$.}
		\label{secrecy}
	\end{figure}

	To analyze the secrecy, we consider yet another system, the system in Figure~\ref{secrecy}.  The corresponding distribution is given by
	\begin{IEEEeqnarray}{rCl}	
	&&	\bar{Q}_{MIU^n\xi(S^n)Z^n}(m,i,u^n,\zeta^n,z^n) 
	\nonumber\\
	&&	= \frac{1}{2^{nR_M}}\frac{1}{2^{nR_I}}   \mathbbm{1}\{{u}^n(i)=u^n\}  P^{\otimes n}_{\nu(S)Z|U}(\zeta^n,z^n |u^n). 
	\end{IEEEeqnarray}
In this idealized secrecy system  and for any choice of the codebook $\mathcal{C}_U$, the pair $(M, \Xi^n)$ is independent of $(U^n,Z^n)$ and $I_{\bar{Q}}(M, \Xi^n; u^n(I) ,Z^n) =0$. 
	
	Notice next that by the generalized superposition soft-covering lemma \cite{Schieler} and by choosing
	\begin{equation}
		R_J > I( V;\nu(S),Z|U),
	\end{equation} the distribution $\bar{Q}_{MIU^n \xi(S^n)Z^n}$ is close to ${Q}_{MIU^n \xi(S^n)Z^n}$, 
	\begin{IEEEeqnarray}{rCl}
&&		\Pr\left[ \bar{Q}_{MIU^n\xi(S^n) Z^n}-  {Q}_{MIU^n \xi(S^n)Z^n}\|_1 \leq e^{-n\gamma} \right]
	\nonumber\\
&&\hspace{3cm} \geq 1 -e^{-n\delta},
	\qquad \forall m \in \mathcal{M},
	\end{IEEEeqnarray}	
	where probability is  over the random choice of the codebook.
	
	By arguments similar to \cite[Appendix D]{Bunin2017} (which use the boundedness of the alphabets), it  also holds that for 
	any pair of codebooks $\mathcal{C}_U$ and $\mathcal{C}_V$ satisfying 
	\begin{equation}
		\left\|P_{MIU^n\xi(S^n) Z^n} -\bar{Q}_{MIU^n\xi(S^n)Z^n}  \right\| \leq e^{-n\gamma} ,
	\end{equation}
	we have  that 
	\begin{equation}
		\left| I_P(M,\xi(S^n);Z^n) - I_{\bar{Q}}(M,\xi(S^n);Z^n)\right | \to 0 \qquad \textnormal{as } n\to \infty.
	\end{equation}
	Combined with $I_{\bar{Q}}(M, \Xi^n; u^n(I) ,Z^n) =0$ this establishes the desired secrecy requirement. 
	
	\section{Conclusions and Discussion}
	
We introduced the concept of securing  state-information  (besides the message) from an external eavesdropper in an ISAC scenario. Notice that such a setup nicely unifies  the models for state-communication \cite{Choudhuri2012noncausal} and state-masking \cite{Merhav2007} into a single problem.  Specifically, we proposed a coding scheme for a state-dependent wiretap channel where the receiver wishes to estimate the state with predefined distortion, and information about this state-sequence has to be kept secret from an external eavesdropper. At hand of a Gaussian example, we numerically show the influence of the various security constraints on our achievable rate-distortion region.
%
%
\clearpage	
\bibliographystyle{ieeetran}
\bibliography{references.bib}
	
	\clearpage
\appendices
	\section{Simplification of the Rate-Constraints for $\nu(S)=$const.}	\label{sec:simplification}
	
		Applying the Fourier-Motzkin elimination algorithm to the rate-constraints \eqref{eq:conditions} and \eqref{eq:constraints} results in the conditions
	\begin{IEEEeqnarray}{rCl}\label{eq:thm:ach_equivocationb}
		R_M&\leq& I(U,V;Y)-I(U,V;S) 
		\\
		R_M&\leq& I(V;Y\mid U)-	I(V;\nu(S),Z\mid U)
		\label{eq2:Th1b}	\\
		R_M&\leq& I(U,V;Y)-I(V;\nu(S),Z\mid U)-I(U;S)\label{eq2:Th2b}
	\end{IEEEeqnarray}

	If $\nu(S)=$const., then similarly to \cite[Appendix~D]{SD_WT} one can show that without loss in optimality one can restrict to random variables $(U,V,X)$ so that $I(U;Y)\geq I(U;S)$, in which case Constraint~\eqref{eq2:Th2b} is not active and thus can be omitted. To show this, fix a triple $(U,V,X)$ so that $I(U;Y)< I(U;S)$. Then, define the new random variables
	\begin{subequations}
		\begin{IEEEeqnarray}{rCl}
			U' &=& (U, \tilde{V})\\
			V'&=& V\\
			X'&= & X,
		\end{IEEEeqnarray}
	\end{subequations}
	where $\tilde{V}$ is the outcome of passing $V$ to a binary erasure channel with erasure probability $\epsilon \in [0,1]$. Also choose, $\epsilon$ so that 
	\begin{IEEEeqnarray}{rCl}\label{eq:cc}
		I(U';Y)= I(U';S),
	\end{IEEEeqnarray}
	which is possible by continuity and  because for $\epsilon=0$ we have $I(U';Y)> I(U';S)$ and for $\epsilon=1$ we have $I(U';Y)< I(U';S)$ by assumption. Notice that by an appropriate change of the reconstruction function $g$,  the new triple $(U',V',X')$ continues to satisfy the distortion constraint \eqref{eq:distortionc}.
	
	Under condition~\eqref{eq:cc}, constraints \eqref{eq2:Th1b} and \eqref{eq2:Th2b} are equivalent, and we shall thus ignore  \eqref{eq2:Th2b}  for the new tuple $(U',V',X')$. Notice further that  when  $(U,V,X)$ is replaced by $(U',V',X')$,  Constraint~\eqref{eq:thm:ach_equivocationb} remains unchanged whereas   constraint  \eqref{eq2:Th1b}  is relaxed because $I(V; Z|U) \geq I(V;Z|U')= I(V';Z'|U')$ due to the Markov chains $U\to U' \to V$  and $(U',V')\to (U,V) \to Z$. We can thus conclude that replacing $(U,V,X)$  by $(U',V',X')$ relaxes Constraints \eqref{eq:thm:ach_equivocationb}--\eqref{eq2:Th2b}, and for the latter triple Constraint \eqref{eq2:Th2b} is inactive. This  proves that for $\nu(S)=$const., Constraints~\eqref{eq:thm:ach_equivocationb}--\eqref{eq2:Th2b} can be replaced by the two constraints
	\begin{IEEEeqnarray}{rCl}
		R_M&\leq& I(U,V;Y)-I(U,V;S) 
		\\
		R_M&\leq& I(V;Y\mid U)-	I(V;Z\mid U)
	\end{IEEEeqnarray}
	and one can restrict to pmfs satisfying $I(U;Y)\geq I(U;S)$.

\end{document}